\documentclass[eqsecnum,preprint,prd,aps,nofootinbib]{revtex4}
\usepackage{amsmath}
\usepackage{psfig}

\begin{document}
\begin{center}
\bibliographystyle{article}
{\Large \textsc{Spherically symmetric ADM gravity with variable $G$ and 
$\Lambda_{c}$}}
\end{center}

\author{Giampiero Esposito$^{1,2}$ \thanks{
Electronic address: giampiero.esposito@na.infn.it} Claudio Rubano$^{2,1}$ 
\thanks{
Electronic address: claudio.rubano@na.infn.it} and Paolo Scudellaro$^{2,1}$ 
\thanks{
Electronic address: scud@na.infn.it}}
\affiliation{${\ }^{1}$Istituto Nazionale di Fisica Nucleare, 
Sezione di Napoli,\\
Complesso Universitario di Monte S. Angelo, Via Cintia, Edificio 6, 80126
Napoli, Italy\\
${\ }^{2}$Dipartimento di Scienze Fisiche, Complesso Universitario di Monte
S. Angelo,\\
Via Cintia, Edificio 6, 80126 Napoli, Italy}

\vspace{0.4cm}
\date{\today}

\begin{abstract}
This paper investigates the Arnowitt--Deser--Misner (hereafter ADM) form of
spherically symmetric gravity with variable Newton parameter $G$ and
cosmological term $\Lambda_{c}$. The
Newton parameter is here treated as a dynamical variable, rather than being
merely an external parameter as in previous work on closely related topics.
The resulting Hamilton equations are obtained; interestingly, a static
solution exists, that reduces to Schwarzschild geometry in the limit 
of constant $G$, describing a
Newton parameter ruled by a nonlinear differential equation in the radial
variable $r$. A remarkable limiting case is the one for which the Newton
parameter obeys an almost linear growth law at large $r$. An exact solution
for $G$ as a function of $r$ is also obtained in the case of vanishing
cosmological constant. Some observational implications of these solutions
are obtained and briefly discussed.
\end{abstract}

\maketitle
\bigskip
\vspace{2cm}

\section{Introduction}

The last ten years have witnessed an encouraging progress in the application
of renormalization-group methods to the nonperturbative renormalization of
Quantum Einstein Gravity \cite{Reut98, Laus05, Nied06, Nied07}. 
In general, field theories
which are nonperturbatively renormalizable are constructed by performing the
limit of infinite ultraviolet cutoff at a nonGaussian renormalization group
fixed point $g_{*i}$ in the space $\left \{ g_{i} \right \}$ of all
dimensionless couplings $g_{i}$ which parametrize a general action
functional. In the case of general relativity, an effective average action
has been built \cite{Reut98}, and such a nonGaussian ultraviolet fixed point
has been found in the case of the Einstein--Hilbert and higher-derivative
truncations \cite{Laus05}.

Several cosmological applications of this framework have also been
considered. For example, in \cite{Weye06} 
(see also \cite{Weye04}) it has been argued that the
resulting scale dependence of the Newton parameter at large distances might
mimic the presence of dark matter at galactic and cosmological scales. On
the other hand, in early work by some of us \cite{Bona04}, we had tried to
build an action functional where the running of the Newton parameter is
ruled by suitable Euler--Lagrange or Hamilton equations, while being
compatible with the renormalization-group flow in the neighbourhood of an
ultraviolet fixed point. For this purpose, one adds to an action of the
Einstein--Hilbert type (but with variable $G$, so that it is brought within
the integrand) two compensating terms such that the action reduces to the
York--Gibbons--Hawking \cite{York72, Gibb77} action for fixed $G$ and 
$\Lambda_{c}$, and takes the same functional form as in the ADM formalism for
general relativity (see Eq. (2.1) below).

The work in Ref. \cite{Bona04} focused on cosmological models with
Friedmann--Lemaitre--Robertson--Walker symmetry, but of course other
symmetries are also relevant in the investigation of the early universe. In
particular, we are here concerned with the requirement of spherical
symmetry, for which a thorough Hamiltonian analysis in general relativity
was performed, for example, in Ref. \cite{Kuch94}. In that case, the
starting point is of course the Schwarzschild line element written in the
curvature coordinates $(T,R)$, i.e. 
\begin{equation}
ds^{2}=-F(R)dT^{2}+F^{-1}(R)dR^{2}+R^{2}(d\theta^{2} +\sin^{2}\theta
d\phi^{2}),  
\label{(1.1)}
\end{equation}
where, in $c=1$ units, $F(R)=1-{\frac{2GM }{R}}$. A space-time foliation is
then introduced, with $R$ and $T$ smooth functions of new independent
variables $r$ and $t$.

At this stage, section 2 generalizes this construction to models with variable
Newton parameter $G$ and cosmological term $\Lambda_{c}$, obtaining the
general Hamilton equations for such models and proving that a static
solution, compatible with the fixed-point hypothesis, actually exists. Section
3 studies the resulting nonlinear differential equation for $G$, while some
phenomenological implications are analyzed in section 4, and concluding
remarks are presented in section 5.

\section{ADM action and Hamilton equations with spherical symmetry and
variable $G, \Lambda_{c}$}

From the analysis in Ref. \cite{Bona04} we know that the ADM action for a
theory of pure gravity where the Newton parameter $G$ and the cosmological
term $\Lambda_{c}$ evolve in space-time as a result of renormalization-group
equations in the early Universe, can be taken to have the form 
\begin{equation}
S={\frac{1}{16 \pi}}\int_{M}\left[{\frac{K_{ij}K^{ij}-K^{2} +{\ }^{(3)}R-2
\Lambda_{c} }{G}}-\mu g^{\rho \sigma} 
{\frac{G_{; \rho} G_{; \sigma} }{G^{3}}}\right]N \sqrt{h}d^{3}x \; dt,  
\label{(2.1)}
\end{equation}
where $K_{ij}$ is the extrinsic curvature tensor of the spacelike
hypersurfaces $\Sigma$ which foliate the space-time 
manifold, while ${\ }^{(3)}R$ is their 
scalar curvature, and $\mu$ is an arbitrary dimensionless parameter.
In other words, it is possible to generalize the standard ADM Lagrangian
and regard $G$ as a dynamical field obeying and Euler--Lagrange equation,
the underlying idea being that all fields occurring in the Lagrangian $L$
should be ruled by $L$ in the first place. This makes it possible to fully
exploit the potentialities of the action principle. For this purpose,
one adds to an action of the Einstein--Hilbert type (but with variable
$G$, so that $G$ is brought within the integrand) two compensating
terms \cite{Bona04} such that the action reduces to the 
York--Gibbons--Hawking \cite{York72, Gibb77} form for fixed $G$ and 
$\Lambda$, and takes the same functional form as the ADM action for
general relativity. Non-vanishing values of $\mu$ in (2.1) 
(cf \cite{Shap95}) ensure that
no primary constraint of vanishing conjugate momentum to $G$ arises.
Its preservation in time (which is necessary because $G$ is here a
dynamical variable, on the same ground of the space-time metric) 
would lead, following the Dirac method \cite{Dira01}, to further
constraints with considerable technical complications.
Such an action is sufficient for the purposes of an ADM analysis, 
which is what we do hereafter, but its
generalization to a form invariant under diffeomorphisms in $4$ 
space-time dimensions remains an open problem. We should also stress that
(2.1) is not just a Brans--Dicke action in the Jordan frame \cite{Fara98},
subject to the identification of ${1\over G}$ with a scalar field $\phi$,
because (2.1), as we said above, results from the definition \cite{Bona04}
\begin{equation}
S \equiv {1\over 16 \pi}\int_{M} \left[{({ }^{(4)}R-2 \Lambda_{c})\over G}
-\mu g^{\rho \sigma}{G_{; \rho}G_{; \sigma}\over G^{3}}\right]
\sqrt{-g}d^{4}x
+{1\over 8\pi}\int_{M}{1\over \sqrt{-g}G} 
\Bigr[(K \sqrt{h})_{,0} -f_{\; ,i}^{i}\Bigr]\sqrt{-g}d^{4}x,
\label{(2.2)}
\end{equation}
where \cite{DeWi67}
\begin{equation}
f^{i} \equiv \sqrt{h}(K N^{i}-h^{ij}N_{,j}).
\label{(2.3)}
\end{equation}
In the formula (2.2), 
the last two terms in the integrand are not total
derivatives since $G$ is variable, and hence the Euler--Lagrange equations
resulting from (2.2) differ from the Brans--Dicke field equations.

On relying upon the work in Ref. \cite{Kuch94} we know that, in a
spherically symmetric space-time, foliated by leaves $\Sigma$ labelled by a
real time parameter $t$, only the radial component $N^{r}$ of the shift
vector survives, and both the lapse $N$ and $N^{r}$ depend only on the $
(t,r) $ variables. Moreover, the three-metric of the leaves reads as \cite
{Kuch94} 
\begin{equation}
ds^{2}=\Lambda^{2}(r,t)dr^{2}+R^{2}(r,t) (d\theta^{2}
+\sin^{2} \theta d\phi^{2}),  
\label{(2.4)}
\end{equation}
where $R$ is the curvature radius of the two-sphere $r = \mathrm{constant}$,
so that the space-time four-metric reads
\begin{equation}
d\sigma^{2}=-\Bigr(N^{2}-(N^{r})^{2}\Bigr)dt^{2}+2N^{r}dr \; dt+ds^{2}.
\label{(2.5)}
\end{equation}
Integration over $\theta$ and $\phi$ in Eq. (2.1) yields therefore
\begin{equation}
S_{\Sigma}[R,\Lambda,G;N,N^{r}]=\int L \; dt,  
\label{(2.6)}
\end{equation}
with Lagrangian (hereafter, following \cite{Kuch94}, dots and primes denote
partial derivatives with respect to $t$ and $r$, respectively) 
\begin{eqnarray}
L&=& \int_{-\infty}^{\infty}dr \biggr[-N^{-1}R ({\dot \Lambda}-(\Lambda
N^{r})')({\dot R}-R' N^{r}) 
-{\frac{1}{2}}N^{-1}\Lambda ({\dot R}-R' N^{r})^{2}  \nonumber \\
&+& N \biggr(-\Lambda^{-1}R
R''+\Lambda^{-2}RR'\Lambda'-{\frac{1}{2}}
\Lambda^{-1}{R'}^{2}+{\frac{1}{2}}\Lambda \biggr)  \nonumber \\
&-& {\frac{N \Lambda R^{2}}{2G}}\Lambda_{c} 
+{\frac{\mu }{4}}{\frac{N \Lambda R^{2}}{G^{3}}} 
\left({\frac{{\dot G}^{2}}{N^{2}}}-2{\frac{N^{r}}{N^{2}}}G'{\dot G} 
+\left(\left({\frac{N^{r}}{N}}\right)^{2}
-{\frac{1}{\Lambda^{2}}}\right) {G^{\prime}}^{2}\right)\biggr].  
\label{(2.7)}
\end{eqnarray}
It should be stressed that we have made a non-trivial step, i.e. the
insertion of spherical-symmetry ansatz into the action before performing the
variations that lead to the field equations. In general relativity, 
spherical reduction in the variational principle leads indeed to the
Schwarzschild solution (see comments below, in between Eqs. (2.38) and 
(2.42)). Mathematicians have realized, by now, what are the
symmetry groups for which the reduction fails (see, for example, the work in
Ref. \cite{Ande00}). For theories with variable $G$, the rigorous proof that
spherical reduction in the variational principle is admissible is an open 
problem, but we will see at the end of section 2 that the resulting 
Hamiltonian constraint is compatible with a space-time metric which, in
the limit of constant $G$, reduces to the Schwarzschild metric. 

At this stage, we might write directly the Euler--Lagrange equations 
resulting from the Lagrangian (2.7). However, since the latter involves
the explicit $r$-integration, the passage to Hamiltonian variables leads 
to a more manageable system of coupled first-order equations 
(see (2.19), (2.20) below).
In the final form of the Hamilton equations one can eliminate the momenta 
and hence recover the desired Euler--Lagrange equations, if necessary,
or rather go the other way round. One may further check the equivalence
of the Hamiltonian and Lagrangian formalism in the relevant case of the
Euler--Lagrange equation for $G$ itself, which is possibly the major 
novelty resulting from the action (2.1). On writing the Lagrangian (2.7) as
$$
L=\int_{-\infty}^{\infty}{\widetilde L}(r,t)dr,
$$
the Euler--Lagrange equation for $G$, i.e.
$$
{d\over dt}{\partial {\widetilde L}\over \partial {\dot G}}
-{\partial {\widetilde L}\over \partial G}=0,
$$
leads to (on setting $N^{r}=0$ for simplicity)
$$
{d\over dt}\left({\mu \Lambda R^{2}\over 2NG^{3}}{\dot G}\right)
=N \left[-{3\over 4}{\mu \Lambda R^{2}\over G^{4}}
{{\dot G}^{2}\over N^{2}}-{\partial \over \partial G}
\left({\Lambda R^{2}\over 2}{\Lambda_{c}\over G}
+{\mu \over 4}{R^{2}\over \Lambda} {{G'}^{2}\over G^{3}}
\right)\right].
$$
But this coincides with the third Hamilton equation (2.28) (see below),
by exploiting 
$$
{\dot G}=N {2G^{3}\over \mu \Lambda R^{2}}p_{G},
$$
which is the third Hamilton equation (2.27) when $N^{r}=0$.

With this understanding we further remark that, by differentiating the ADM
action (2.3) with respect to the velocities ${\dot \Lambda},{\dot R}$ 
and ${\dot G}$, we obtain the momenta (cf. Ref. \cite{Kuch94}) 
\begin{equation}
p_{\Lambda}=-N^{-1}R ({\dot R}-R' N^{r}),  
\label{(2.8)}
\end{equation}
\begin{equation}
p_{R}=-N^{-1}\Bigr[\Lambda ({\dot R}-R' N^{r}) +R ({\dot \Lambda}
-(\Lambda N^{r})') \Bigr],  
\label{(2.9)}
\end{equation}
\begin{equation}
p_{G}={\frac{\mu }{2}}{\frac{\Lambda R^{2}}{N G^{3}}}({\dot G}
-G' N^{r}).  
\label{(2.10)}
\end{equation}
Equations (2.8)--(2.10) can be inverted for the velocities, i.e. 
\begin{equation}
{\dot \Lambda}=-N R^{-2}(R p_{R}-\Lambda p_{\Lambda}) 
+(\Lambda N^{r})',  
\label{(2.11)}
\end{equation}
\begin{equation}
{\dot R}=-NR^{-1}p_{\Lambda}+R' N^{r},  
\label{(2.12)}
\end{equation}
\begin{equation}
{\dot G}={\frac{2}{\mu}}{\frac{NG^{3}}{\Lambda R^{2}}}p_{G}+G' N^{r}.
\label{(2.13)}
\end{equation}

The ADM action (2.6) can be cast into the canonical form by the Legendre
transform 
\begin{equation}
S_{\Sigma}[\Lambda,R,G,p_{\Lambda},p_{R},p_{G};N,N^{r}] =\int dt
\int_{-\infty}^{\infty}dr \Bigr(p_{\Lambda} {\dot \Lambda} +p_{R}{\dot R}
+p_{G}{\dot G}-NH-N^{r}H_{r}\Bigr).  
\label{(2.14)}
\end{equation}
The insertion of Eqs. (2.11)--(2.13) into (2.7) and (2.14) leads to two
equivalent expressions of the Lagrangian, so that the functions multiplying
lapse and shift therein must be equal. Hence we find 
(cf. Ref. \cite{Kuch94}) 
\begin{eqnarray}
H(r,t)&=& -{\frac{p_{R}p_{\Lambda}}{R}} +{\frac{1}{2}}{\frac{\Lambda
p_{\Lambda}^{2}}{R^{2}}} +{\frac{G^{3}}{\mu \Lambda R^{2}}}p_{G}^{2} 
+{\frac{R R''}{\Lambda}}
-{\frac{R R' \Lambda'}{\Lambda^{2}}} 
+{\frac{{R'}^{2}}{2 \Lambda}}  \nonumber \\
&-& {\frac{1}{2}}\Lambda +{\frac{\Lambda R^{2} }{2G}}\Lambda_{c}
+{\frac{\mu}{4}} {\frac{{G'}^{2}R^{2}}{\Lambda G^{3}}},  
\label{(2.15)}
\end{eqnarray}
\begin{equation}
H_{r}(r,t)=R'p_{R}-\Lambda p_{\Lambda}'+G' p_{G}.
\label{(2.16)}
\end{equation}
In the course of deriving Eq. (2.16), we have found in 
$p_{\Lambda}{\dot \Lambda}+p_{R}{\dot R}$ from Eq. (2.14) a term 
\begin{equation}
(N^{r})'\Lambda p_{\Lambda}=-{\frac{\partial }{\partial r}}
(N^{r}\Lambda p_{\Lambda})-N^{r}\Lambda' p_{\Lambda} 
-N^{r}\Lambda p_{\Lambda}'.  
\label{(2.17)}
\end{equation}
Thus, Eq. (2.16) holds because, upon $r$-integration, the first term on the
right-hand side of Eq. (2.17) gives vanishing contribution subject to the
fall-off conditions in Sec. IIIC of Ref. \cite{Kuch94}. The second term on
the r.h.s. of Eq. (2.17) is cancelled exactly in the integrand of Eq.
(2.14), while the third term on the r.h.s. of Eq. (2.17) leads to 
$-\Lambda p_{\Lambda}'$ in Eq. (2.16). The effective Hamiltonian now reads 
\begin{equation}
{\widetilde H}(t)=\int_{-\infty}^{\infty}[NH+N^{r}H_{r}+\nu_{N}\pi_{N}
+\nu^{i}\pi_{i}](\rho,t)d\rho,  
\label{(2.18)}
\end{equation}
where $\pi_{N}$ and $\pi_{i}$ are the primary constraints which occur
because the ADM Lagrangian (2.7) is independent of time derivatives of lapse
and shift. The general Hamilton equations are therefore 
\begin{equation}
{\frac{d}{dt}}Q(r,t)= \left \{ Q(r,t),{\widetilde H}(t) \right \}, \;
Q=\Lambda,R,G,  
\label{(2.19)}
\end{equation}
\begin{equation}
{\frac{d}{dt}}p_{Q}(r,t)= \left \{ p_{Q}(r,t),{\widetilde H}(t) \right \},
\; p_{Q}=p_{\Lambda},p_{R},p_{G},  
\label{(2.20)}
\end{equation}
to be solved for given initial conditions $\Lambda(0), R(0), G(0),
p_{\Lambda}(0), p_{R}(0), p_{G}(0)$ satisfying the constraint equations 
\begin{equation}
H \approx 0, \; H_{r} \approx 0,  
\label{(2.21)}
\end{equation}
where the weak-equality symbol $\approx$ denotes equations which only hold
on the constraint manifold \cite{Dira01}.

We now exploit the relations 
\begin{equation}
{\frac{d}{dt}} Q(r,t) \approx \int_{-\infty}^{\infty} \Bigr[N(\rho,t) \left
\{ Q(r,t),H(\rho,t) \right \} +N^{r}(\rho,t) \left \{
Q(r,t),H_{r}(\rho,t)\right \} \Bigr]d\rho,  
\label{(2.22)}
\end{equation}
\begin{equation}
{\frac{d}{dt}}p_{Q}(r,t) \approx \int_{-\infty}^{\infty}\Bigr[N(\rho,t)
\left \{ p_{Q}(r,t),H(\rho,t) \right \} +N^{r}(\rho,t) \left \{
p_{Q}(r,t),H_{r}(\rho,t) \right \} \Bigr]d\rho,  
\label{(2.23)}
\end{equation}
\begin{equation}
\left \{ Q(r,t),H_{d}(\rho,t) \right \} ={\frac{\partial H_{d} }{\partial
p_{Q}}}(\rho,t)\delta(r,\rho), \; H_{d}=H,H_{r},  
\label{(2.24)}
\end{equation}
\begin{equation}
\left \{ p_{Q}(r,t),H_{d}(\rho,t) \right \} 
=-{\frac{\partial H_{d}}{\partial Q}}(\rho,t) 
\delta(r,\rho), \; H_{d}=H,H_{r},  
\label{(2.25)}
\end{equation}
and define the vector field 
\begin{equation}
D_{tr} \equiv {\frac{\partial }{\partial t}} 
-N^{r}{\frac{\partial }{\partial r}}  
\label{(2.26)}
\end{equation}
to find, for all values of lapse and shift, the general Hamilton equations 
\begin{equation}
D_{tr}\Lambda \approx N U_{1}, \; D_{tr}R \approx N U_{2}, \; D_{tr}G
\approx N U_{3},  
\label{(2.27)}
\end{equation}
\begin{equation}
D_{tr}p_{\Lambda} \approx N V_{1}, \; D_{tr}p_{R} \approx N V_{2}, \;
D_{tr}p_{G} \approx N V_{3},  
\label{(2.28)}
\end{equation}
having set 
\begin{equation}
U_{1} \equiv -{\frac{p_{R}}{R}}+{\frac{\Lambda p_{\Lambda}}{R^{2}}},
\label{(2.29)}
\end{equation}
\begin{equation}
U_{2} \equiv -{\frac{p_{\Lambda}}{R}},  
\label{(2.30)}
\end{equation}
\begin{equation}
U_{3} \equiv {\frac{2G^{3}}{\mu \Lambda R^{2}}}p_{G},  
\label{(2.31)}
\end{equation}
\begin{equation}
V_{1} \equiv -{\frac{p_{\Lambda}^{2}}{2R^{2}}} +{\frac{G^{3}}{\mu
\Lambda^{2}R^{2}}}p_{G}^{2} 
-{\frac{\partial }{\partial \Lambda}}\left({\frac{RR''}{\Lambda}} 
-{\frac{RR'\Lambda'}{\Lambda^{2}}}
+{\frac{{R'}^{2}}{2 \Lambda}} \right) 
+ {\frac{1}{2}}-{\frac{R^{2}}{2G}}\Lambda_{c} 
+{\frac{\mu }{4}}{\frac{{G'}^{2}R^{2}}{\Lambda^{2}G^{3}}},  
\label{(2.32)}
\end{equation}
\begin{equation}
V_{2} \equiv -{\frac{p_{R}p_{\Lambda}}{R^{2}}}
+{\frac{\Lambda p_{\Lambda}^{2}}{R^{3}}} 
+{\frac{2G^{3}}{\mu \Lambda R^{3}}}p_{G}^{2} 
-{\frac{\partial}{\partial R}}\left({\frac{R R''}{\Lambda}} 
-{\frac{RR' \Lambda'}{\Lambda^{2}}}
+{\frac{{R'}^{2}}{2\Lambda}}\right) 
- {\frac{\Lambda R }{G}} \Lambda_{c} 
-{\frac{\mu }{2}}{\frac{{G'}^{2}R }{\Lambda G^{3}}},  
\label{(2.33)}
\end{equation}
\begin{equation}
V_{3} \equiv -{\frac{3G^{2}}{\mu \Lambda R^{2}}}p_{G}^{2} 
-{\frac{\partial }{\partial G}} \left({\frac{\Lambda R^{2}}{2}}
{\frac{\Lambda_{c}}{G}} +{\frac{\mu }{4}}
{\frac{R^{2}}{\Lambda}} {\frac{{G^{'}}^{2}}{G^{3}}}\right).
\label{(2.34)}
\end{equation}
The six equations (2.27) and (2.28) should be studied, for given initial
conditions, jointly with the two constraint equations (2.21) and with the
ADM relations for lapse and shift in the $(t,r,\theta,\phi)$ coordinates,
i.e. \cite{Kuch94} 
\begin{equation}
\Lambda \equiv \sqrt{-F{T'}^{2}+F^{-1}{R'}^{2}},
\label{(2.35)}
\end{equation}
\begin{equation}
N={\frac{R'{\dot T}-T' {\dot R}}{\Lambda}},  
\label{(2.36)}
\end{equation}
\begin{equation}
N^{r}={\frac{-FT'{\dot T}+F^{-1}R'{\dot R}}{\Lambda^{2}}}.
\label{(2.37)}
\end{equation}

It is reassuring to note that such equations make it possible to recover the
Schwarzschild solution in general relativity. For this purpose, it is enough
to choose the foliation defined by 
\begin{equation}
R(r,t)=r, \; T(r,t)=t,  
\label{(2.38)}
\end{equation}
for which $N^{r}=0$ and hence $p_{\Lambda} \approx 0$, $p_{R} \approx 0$,
with Hamiltonian constraint $H \approx 0$ reducing to (from Eq. (2.15)) 
\begin{equation}
{\frac{R R' \Lambda'}{\Lambda^{2}}} 
\approx {\frac{1}{2\Lambda}}(1-\Lambda^{2}).  
\label{(2.39)}
\end{equation}
Such an equation is solved by $R=r$ and 
\begin{equation}
\Lambda=F^{-1/2}=\left(1-{\frac{2GM }{r}}\right)^{-1/2}.  
\label{(2.40)}
\end{equation}
Moreover, the weak equation (2.39) can then be used to cast Eq. (2.32) in
the form 
\begin{eqnarray}
V_{1} & \approx & -{\frac{2}{\Lambda}}
{\frac{R R' \Lambda'}{\Lambda^{2}}} 
+{\frac{R R'}{\Lambda^{2}}}
{\frac{\partial \Lambda'}{\partial \Lambda}} 
+{\frac{{R'}^{2}}{2\Lambda^{2}}}+{\frac{1}{2}}  \nonumber \\
& \approx & -{\frac{2}{\Lambda}}\left({\frac{1}{2\Lambda}} 
-{\frac{\Lambda }{2}}\right)
+{\frac{1}{2 \Lambda^{2}}} {\frac{\partial}{\partial \Lambda}}
(\Lambda-\Lambda^{3}) +{\frac{1}{2\Lambda^{2}}}+{\frac{1}{2}} \approx 0.
\label{(2.41)}
\end{eqnarray}
Along the same lines, we obtain the weak equation 
\begin{eqnarray}
V_{2} & \approx & -{\frac{\partial }{\partial R}} \left({\frac{R
R''}{\Lambda}}
-{\frac{R R'\Lambda'}{\Lambda^{2}}} 
+{\frac{{R'}^{2}}{2\Lambda}}\right)  \nonumber \\
& \approx & -{\frac{1}{\Lambda}}{\frac{\partial}{\partial R}} 
\left[{\frac{\partial}
{\partial R}}(RR')-{R'}^{2} -{\frac{1}{2}}
(1-\Lambda^{2})+{\frac{{R'}^{2}}{2}}\right]  \nonumber \\
& \approx & -{\frac{1}{\Lambda}}{\frac{\partial^{2}}{\partial R^{2}}}
(RR')+{\frac{1}{2\Lambda}}{\frac{\partial}
{\partial R}}({R'}^{2}) \approx 0.  
\label{(2.42)}
\end{eqnarray}
Our remark agrees with the findings in \cite{Berg72}, but our analysis
offers the advantage of not having to eliminate $R$ and $p_{R}$ from the
Hamiltonian analysis, which is important when $G$ and $\Lambda_{c}$ are
allowed to vary.

In the latter case, we can solve the general Hamilton equations expressed by
(2.27)--(2.34) when the space-time foliation is again given by (2.38) with
vanishing shift and $p_{\Lambda} \approx 0, \; p_{R} \approx 0, \; p_{G}
\approx 0$. In this static case, where only the spatial gradient of $G$ is
nonvanishing, we {\it assume} a fixed-point relation as in \cite{Bona04}: 
\begin{equation}
\Lambda_{c}G=\Lambda_{c}(r)G(r)=\mathrm{constant}=k,  
\label{(2.43)}
\end{equation}
so that the Hamiltonian constraint $H \approx 0$ yields now 
\begin{equation}
{\frac{RR'\Lambda'}{\Lambda^{2}}} 
\approx {\frac{1}{2\Lambda}} (1-\Lambda^{2})
+{\frac{k \Lambda R^{2}}{2G^{2}}} 
+{\frac{\mu }{4}}{\frac{{G'}^{2}R^{2}}{\Lambda G^{3}}}.  
\label{(2.44)}
\end{equation}
The functions $V_{1},V_{2}$ and $V_{3}$ defined in (2.32)--(2.34) are again
weakly vanishing, since, by virtue of (2.38), (2.43) and (2.44), 
\begin{eqnarray}
V_{1} & \approx & -{\frac{2}{\Lambda}} \left[{\frac{1}{2\Lambda}}
(1-\Lambda^{2})+{\frac{k \Lambda R^{2} }{2G^{2}}} 
+{\frac{\mu }{4}}{\frac{{G'}^{2}R^{2}}
{\Lambda G^{3}}}\right]  \nonumber \\
&+& {\frac{1}{\Lambda^{2}}}{\frac{\partial }{\partial \Lambda}} 
\left[{\frac{\Lambda }{2}}(1-\Lambda^{2}) 
+{\frac{k \Lambda^{3}R^{2}}{2G^{2}}} 
+{\frac{\mu }{4}}{\frac{\Lambda {G'}^{2}R^{2}}{G^{3}}}\right]  \nonumber \\
&+& {\frac{{R'}^{2}}{2\Lambda^{2}}}+{\frac{1}{2}}
-{\frac{kR^{2}}{2G^{2}}} +{\frac{\mu}{4}}
{\frac{{G'}^{2}R^{2}}{\Lambda^{2}G^{3}}} \approx 0,
\label{(2.45)}
\end{eqnarray}
\begin{eqnarray}
V_{2} & \approx & {\frac{\partial}{\partial R}} 
\left[{\frac{1}{2\Lambda}}
(1-\Lambda^{2})+{\frac{k \Lambda R^{2}}{2G^{2}}} 
+{\frac{\mu }{4}}{\frac{{G'}^{2}R^{2}}{\Lambda G^{3}}}\right] 
-{\frac{k \Lambda R}{G^{2}}} 
-{\frac{\mu }{2}}{\frac{{G'}^{2}R }
{\Lambda G^{3}}}  \nonumber \\
& \approx & 0,  
\label{(2.46)}
\end{eqnarray}
\begin{eqnarray}
V_{3} & \approx & -{\frac{\partial}{\partial G}} 
\left({\frac{\Lambda R^{2}}{2}}{\frac{\Lambda_{c}}{G}} 
+{\frac{\mu }{4}}{\frac{R^{2}}{\Lambda}} 
{\frac{{G'}^{2}}{G^{3}}}\right)  \nonumber \\
& \approx & 
-{\frac{\partial }{\partial G}} \left({\frac{RR' \Lambda'}{\Lambda^{2}}} 
-{\frac{1}{2\Lambda}}(1-\Lambda^{2})\right) \approx 0.  
\label{(2.47)}
\end{eqnarray}

If no infrared fixed-point relation such as (2.43) can be assumed, Eqs.
(2.44)--(2.47) still hold provided that one replaces $k$ therein by
the product $\Lambda_{c}G$. For example, the Hamiltonian constraint (2.44)
takes the form 
\begin{equation}
{\frac{RR'\Lambda'}{\Lambda^{2}}} 
\approx {\frac{1}{2\Lambda}} (1-\Lambda^{2})
+{\frac{\Lambda R^{2}}{2G}}\Lambda_{c} 
+{\frac{\mu }{4}}{\frac{{G'}^{2}R^{2}}{\Lambda G^{3}}}.  
\label{(2.48)}
\end{equation}
Moreover, it is always true that $\Lambda_{c}$
and $G$ are not independent variables but are functionally related. This 
is clearly proved in the Hamiltonian framework advocated in our paper. Suppose
in fact that $\Lambda_{c}$ were an independent dynamical variable. The primary 
constraint of vanishing momentum conjugate to $\Lambda_{c}$ would then be
preserved in time, from (2.1), provided that 
either the lapse function or the 
determinant of the induced three-metric vanishes, leading therefore to a
complete `collapse' of the ADM geometry. 

\section{Nonlinear differential equation for the Newton parameter}

We now also assume that the departure from general relativity is not so
severe, so that the $\Lambda$ function keeps its functional dependence
on $G(r)$, at least approximately. More precisely, since $\Lambda$
should reduce to (2.40) in the case of constant $G$, it can only differ
from (2.40) by terms involving explicitly the gradient of $G$. A good 
`a posteriori' check of any approximate solution for $G$ is therefore
whether it has a gradient with negligible effects. We thus 
insert into Eq. (2.44) the form (2.40) of the $\Lambda $ function,
with $G$ taken to depend on $r$ only, and we obtain eventually 
the nonlinear differential equation 
\begin{equation}
A(r){G'}^{2}(r)+{G^{2}(r)}G'(r)-{B(r)=0},  
\label{(3.1)}
\end{equation}
where (hereafter, $G_{n}$ is the Newton parameter on solar system scale)
\begin{equation}
A(r)\equiv {\frac{\mu }{2}}r
\left(1-{\frac{rc^{2}}{2MG(r)}}\right) G_{n}<0,
\label{(3.2)}
\end{equation}
\begin{equation}
B(r)\equiv {\frac{k}{2M}}r^{2}{c}^{2}G_{n}>0,  
\label{(3.3)}
\end{equation}
and we have restored the physical units, since we want to make
estimates on real situations.
The statement $A<0$ is true only for $\mu >0$ and $r$ sufficienly large,
which is surely true for normal astrophysical objects, like sun and
galaxies. The case of the immediate neighbourhood
of a black hole horizon is more involved and goes beyond the aims of
the present paper (cf \cite{Bona00}). 

We therefore see that Eq. (\ref{(3.1)}) admits always two distinct positive
values for $G'(r)$. This means that in any case $G(r)$ is
monotonically increasing with $r$. We shall see in a moment that this can be
reconciled with the requirement that the metric should be of Minkowski type
at infinity.  Another problem is posed by the two disjoint solutions. We
assume however that $\mu$ is sufficiently small to get the first term of
Eq. (\ref{(3.1)}) negligible in our case. We have thus to treat a much
simpler equation (a more accurate treatment and justification of this
assumption is given in the next section), i.e.
\begin{equation}
G^{2}dG={\frac{kc^{2}G_{n}}{2M}}r^{2}dr,  
\label{(3.4)}
\end{equation}
which leads to the growth of the Newton parameter according to 
\begin{equation}
G(r)=G_{n}\left( 1+{\frac{k{c}^{2}}{2MG_{n}^{2}}}r^{3}\right) ^{\frac{1}{3}},
\label{(3.5)}
\end{equation}
where we have set the integration constant, in order to obtain a correction
to Newton's law, as desired.

At large $r$, the Newton parameter obeys therefore the approximately linear
relation 
\begin{equation}
G(r)\sim G_{n}\left( {\frac{kc^{2}}{2MG_{n}^{2}}}\right) ^{\frac{1}{3}}r.
\label{(3.6)}
\end{equation}
Therefore, reverting to Eq. (\ref{(2.37)}), we see that the 
function $F=(1-2GM/rc^{2})$ tends asymptotically to the constant
\begin{equation}
F=1-\left( \frac{4kMG_{n}}{c^{4}}\right) ^{1/3},
\label{(3.7)}
\end{equation}
so that, by rescaling appropriately distance and time, we may obtain 
flat space.

A very interesting feature of Eq. (\ref{(3.6)}) is that it gives just the
correction necessary to obtain perfectly flat rotation curves of galaxies.
Let us indeed rewrite it as
\begin{equation}
G=G_{n}\left( 1+\alpha r_{g}^{3}M_{g}^{-1}\rho _{g}^{3}\right) ^{1/3},
\label{(3.8)}
\end{equation}
where $M_{g}$ is the galaxy mass, $\alpha =\frac{kc^{2}}{2G_{n}^{2}}$ 
and $\rho _{g}=r/r_{g}$ 
is the distance rescaled according with a typical galaxy
length $r_{g}$. We see that the correction is effective 
at say $\rho_{g}\simeq 1.2$, 
if we take $\alpha r_{g}^{3}M_{g}^{-1}\simeq 0.5$.
At the solar system scale we get ($\rho _{s}=r/r_{s}\simeq 1$)
\begin{equation}
G=G_{n}\left( 1+\alpha r_{s}^{3}M_{s}^{-1}\rho _{s}^{3}\right)
^{1/3}=G_{n}\left( 1+\frac{r_{s}^{3}M_{s}^{-1}}
{r_{g}^{3}M_{g}^{-1}}\rho_{s}^{3}\right) ^{1/3}\simeq 
G_{n}\left( 1+10^{-16}\rho _{s}^{3}\right)^{1/3},
\label{(3.9)}
\end{equation}
and the correction is absolutely irrelevant.

It is also interesting to note that, at large $r$, we get 
for the radial velocities of galaxies the relation
\begin{equation}
v=\sqrt{\frac{GM}{r}}\propto M^{1/3},
\label{(3.10)}
\end{equation}
where the proportionality constant is equal to $(kc^{2}G_{n}/2)^{1/6}$.
Now, the usual theoretical expression for the 
Tully--Fisher relation is $v\propto M^{1/4}$, 
and is computed with the usual Keplerian law for
velocities. On the other hand, we have the observational relation 
\begin{equation}
{\rm Mag}=-7.68\log _{10}\left( \frac{2}{\sin i}v\right) -\log_{10}q,
\label{(3.11)}
\end{equation}
where ${\rm Mag}$ is the absolute magnitude, 
$i$ is the visual angle of the galaxy
and $q$ is a number which depends on the optical band  
\cite {Giov97}. The first coefficient, which
is the only relevant one for our purpose, has small dependence on the
band. If we consider also the theoretical definition of magnitude, i.e.
\begin{equation}
{\rm Mag}=-2.5\log _{10}L+\log _{10}a,
\label{(3.12)}
\end{equation}
where $L$ is the absolute luminosity, assumed proportional to 
the mass, and $a$ is again dependent on the band, but irrelevant, 
we obtain eventually 
\begin{equation}
v\propto M^{2.5/7.68}=M^{0.325}\simeq M^{1/3},
\label{(3.13)}
\end{equation}
so that we obtain a striking agreement with our Eq. (3.10), unlike the
work in \cite{Weye06}, where no agreement with the empirical 
Tully--Fisher relation is found (see comments in the last paragraph 
of section 4 therein). The reason for this improvement lies in the fact
that, in our case, the correction (3.5) to the Newtonian $G$ is not 
parametrized only by universal constants, but by the mass of the 
gravitating source of the field. What is instead depending only on
universal constants is the proportionality coefficient $v/M^{1/3}$
in (3.10).

\section{Qualitative analysis of the equation for $G$}

Let us now consider again Eq. (\ref{(3.1)}) and show that indeed its
replacement with the much simpler Eq. (\ref{(3.4)}) is justified. 
First, let us point out that, for any 
astrophysical object different from a black hole we may write safely
\begin{equation}
A(r)\equiv -{\frac{\mu r^{2}c^{2}}{4MG(r)}} G_{n}.
\label{(4.1)}
\end{equation}

Then Eq. (\ref{(3.1)}) may be rewritten, in $c=1$ units, as
\begin{equation}
G'(r)=\frac{G^{3}(r)
\pm \sqrt{G^{6}(r)-2\alpha \beta r^{4}G(r)}}{\alpha r^{2}},  
\label{(4.2)}
\end{equation}
where $\alpha \equiv \mu G_{n}^{2}/r_{s}$, 
$\beta \equiv kG_{n}^{2}/r_{s}$, and $r_{s}=2G_{n}M$ 
is the Schwarzschild radius of the object.

We are considering only one of the two equations generated by 
Eq. (\ref{(3.1)}), 
the study of the other being made along the same lines. We see that we may
reduce the number of relevant parameters to only two.
This equation can be exactly solved in the cases $\alpha =0$ and $\beta =0$.
The first corresponds to the solution examined in the previous section,
which we rewrite here as
\begin{equation}
G(r)=\beta ^{1/3}(r^{3}+3w)^{1/3},
\label{(4.3)}
\end{equation}
where $w$ is an integration constant.

The second case is possibly even more interesting, since it corresponds to
chosing $\Lambda _{c}=0$, which is closer to the Schwarzschild 
geometry. There is then 
a trivial solution $G={\rm constant}$, as well as 
\begin{equation}
G(r)=\sqrt{\frac{\alpha r}{4+2 \chi r}} \; {\rm at} \;
r >> r_{s},
\label{(4.4)}
\end{equation}
where $\chi$ is the integration constant. 
We see that this solution also tends asymptotically to a constant,
and hence satisfies the consistency check stated at the beginnin
of section 3. If 
this regime is reached sufficiently late, an emulation of dark matter 
might be obtained again on taking a linearized approximation
of (4.4) (cf. comments in section 5). 
Everything depends on the values of the parameters and
confrontation with observations.

Let us now show that the first solution dominates at large $r$. For this
purpose, let us consider the ratio of the first to the last term in 
Eq. (\ref{(3.1)}), i.e.
\begin{equation}
\delta =\frac{\alpha {G'}^{2}(r)}{2 \beta G(r)}.
\label{(4.5)}
\end{equation}
It is clear that our approximate treatment will be good as long as $\delta
<<1$. We substitute in $\delta$ the approximate solution and obtain
\begin{equation}
\delta _{1}=\alpha \beta^{-{2/3}} \frac{r^{4}}{2(r^{3}+3w)^{5/3}}.
\label{(4.6)}
\end{equation}
The behaviour of this function is independent of $\alpha $ and $\beta $. It
tends asymptotically to zero and has a maximum 
at $r_{\rm m}\simeq 2.29w^{1/3}$.
Therefore, provided we start the integration 
at $r_{\rm start}>r_{\rm m}$, if the
approximation is valid there, it will be increasingly accurate 
as $r$ gets larger. 

On the other hand, let us suppose that at $r_{\rm start}$ the opposite occurs,
and $\delta (r_{\rm start})>>1$. 
We may thus substitute the other solution, finding
\begin{equation}
\delta _{2}=\frac{2 \alpha ^{3/2}}{\beta r^{3/2}(4+2 \chi r)^{5/2}}.
\label{(4.7)}
\end{equation}
We see that again the behaviour of $\delta _{2}$ is independent of $\alpha$
and $\beta$, but (which is most important) we 
always have that $\delta_{2}^{'}<0$. 
Therefore, even if at the beginning $\delta(r_{start})>>1$, 
as $r$ increases the condition is reversed and we may say
(approximately) that, when $\delta _{2}=1$, we may switch off this solution
and revert to the first one, which prevails asymptotically. The
intermediate situation is of course somewhat delicate, but a numerical
analysis, made on the full equation, with suitable choice of the parameters
involved, confirms these statements. 

\section{Concluding remarks}

In the first part of our paper, we have extended the Hamiltonian analysis
of spherically symmetric gravity \cite{Berg72, Kuch94} to the case of
variable Newton parameter and variable cosmological term, obtaining 
eventually the non-linear differential equation (3.1) for $G(r)$, under
the non-trivial assumption that Eq. (2.40) can be taken to hold. We have 
then shown that {\it the treatment of $\Lambda _{c}$ and $G$ as dynamical
variables, together with the fixed-point condition, gives encouraging chances
of emulating the presence of dark matter in long-range gravitational
interactions}, at least at galactic scale. Several open problems should be
now studied, i.e.
\vskip 0.3cm
\noindent
(i) The legitimacy of the fixed-point assumption.
\vskip 0.3cm
\noindent 
(ii) The validity of Eq. (2.40) when $G$ depends on $r$.
\vskip 0.3cm
\noindent
(iii) The detailed numerical proof that also our solution (4.4) with
vanishing cosmological constant can fit the flat rotation curves of
galaxies. 
\vskip 0.3cm
\noindent
(iv) Can our Hamiltonian approach make it possible to study weak-lensing
observations, that are recently found to provide a direct empirical
proof of the existence of dark matter? \cite{Clow06}.
\vskip 0.3cm
\noindent
(v) How to perform the Hamiltonian analysis 
with variable $G$ in the small-$r$ region which is relevant for
black-hole physics? 

The main source of future developments is the confrontation of our theoretical
results with observational data. This is not a simple task, because a galaxy
is not a pointlike source. Therefore a separate paper is in order
on this topic as well as the other open problems listed above. 
Anyway, since the free parameters of our theory, $\alpha$
and $\beta$, depend on $\mu $ and $k$, which are completely undetermined
at the moment, we hope to be able to obtain 
reliable numerical solutions (see Figs. 1 and 2), appropriate
for comparison with observations.

\begin{figure}[!h]
\centerline{\hbox{\psfig{figure=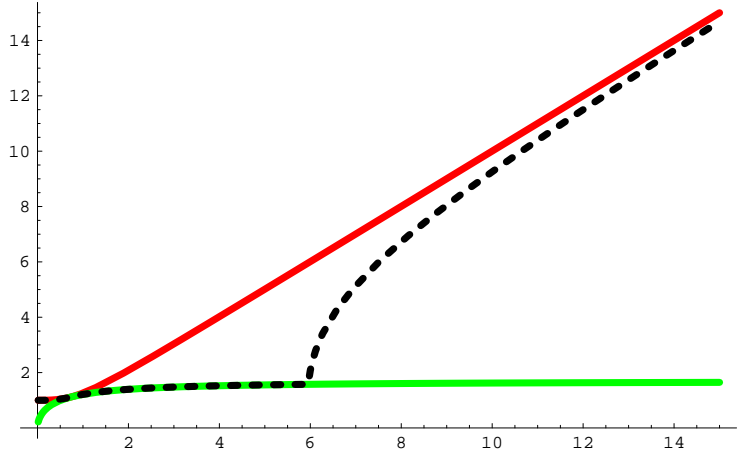,width=0.6\textwidth}}}
\caption{The two approximate solutions of Eq. (4.2) and a rough attempt 
to join them appropriately. The red curve is obtained in the case
$\alpha=0$, the green is for $\beta=0$, so that the red is appropriate
when $\delta <<1$ and the green when $\delta >>1$. The black dashed is
obtained by using the first solution for $\delta <1$, which happens at
the beginning and at the end, and the second in the other case. The
discontinuity in the derivative results from the rough procedure. This
curve should be compared with the ones of Fig. 2, obtained numerically
from the full equation.} 
\end{figure}

\begin{figure}[!h]
\centerline{\hbox{\psfig{figure=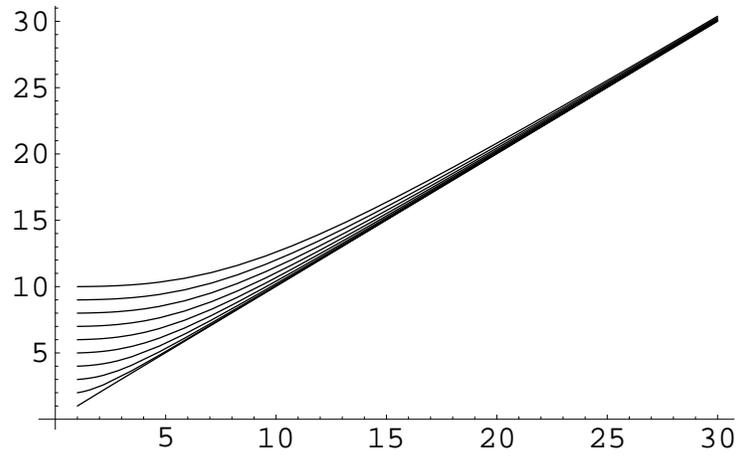,width=0.6\textwidth}}}
\caption{Numerical integration of the full Eq. (4.2) when 
$\alpha=0.1,\beta=1,r_{\rm start}=1$, for various values of
$G_{\rm start}$. Units are arbitrary. Comparison with Fig. 1 shows
that the qualitative analysis is correct.}
\end{figure} 

\acknowledgments The authors are indebted to Alfio Bonanno,
Karel Kuchar and Nelson Pinto Neto for
correspondence, and to the INFN for financial support.


\begin{references}
\bibitem{Reut98}
Reuter M 1998 {\it Phys. Rev.} D {\bf 57} 971 
\bibitem{Laus05}
Lauscher O and Reuter M 2005 hep-th/0511260.
\bibitem{Nied06}
Niedermaier M and Reuter M 2006 {\it Liv. Rev. Rel.} {\bf 5}
\bibitem{Nied07}
Niedermaier M 2007 {\it Class. Quantum Grav.} {\bf 24} R171
\bibitem{Weye06}
Reuter M and Weyer H 2006 {\it Int. J. Mod. Phys.} D {\bf 15} 2011 
\bibitem{Weye04}
Reuter M and Weyer H 2004 {\it Phys. Rev.} D {\bf 69} 104022
\bibitem{Bona04}
Bonanno A, Esposito G and Rubano C 2004 
{\it Class. Quantum Grav.} {\bf 21} 5005 
\bibitem{York72}
York J 1972 {\it Phys. Rev. Lett.} {\bf 28} 1082 
\bibitem{Gibb77}
Gibbons G W and Hawking S W, {\it Phys. Rev.} D {\bf 15} 2752 
\bibitem{Kuch94}
Kuchar K V 1994 {\it Phys. Rev.} D {\bf 50} 3961
\bibitem{Shap95}
Shapiro I L and Takata A 1995 {\it Phys. Rev.} D {\bf 52} 2162 
\bibitem{Dira01}
Dirac P A M 2001 {\it Lectures on Quantum Mechanics} (New York: Dover).
\bibitem{Fara98}
Faraoni V, Gunzig E and Nardone P 1998 {\it Fund. Cosmic Phys.}
{\bf 20} 121 (gr-qc/9811047)
\bibitem{DeWi67}
DeWitt B S 1967 {\it Phys. Rev.} {\bf 160} 1113
\bibitem{Ande00}
Anderson I M, Fels M E and Torre C G 2000 
{\it Commun. Math. Phys.} {\bf 212} 653 
\bibitem{Berg72}
Berger B K, Chitre D M, Moncrief V E and Nutku Y 1972 
{\it Phys. Rev.} D {\bf 5} 2467 
\bibitem{Bona00}
Bonanno A and Reuter M 2000 {\it Phys. Rev.} D {\bf 62} 043008 
\bibitem{Giov97}
Giovanelli R, Haynes M P, Herter T, Vogt N P, Wegner G, Salzer J J,
Da Costa L N and Freudling W 1997 {\it Astron. J.} {\bf 113} 22
\bibitem{Clow06}
Clowe D, Bradac M, Gonzalez A H, Markevitch M, Randall S W, Jones C
and Zaritsky D ``A direct empirical proof of the existence of dark
matter'' (astro-ph/0608407).

\end{references}
\end{document}